# Image denosing in underwater acoustic noise using discrete wavelet transform with different noise level estimation

**Yasin Yousif Al-Aboosi, Radhi Sehen Issa, Ali khalid Jassim**

**Faculty of Engineering, University of Mustansiriyah, Iraq**

**ABSTRACT**

In many applications, Image de-noising and improvement represent essential processes in presence of colored noise such that in underwater. Power spectral density of the noise is changeable within a definite frequency range, and autocorrelation noise function is does not like delta function. So, noise in underwater is characterized as colored noise. In this paper, a novel image de-noising method is proposed using multi-level noise power estimation in discrete wavelet transform with different basis functions. Peak signal to noise ratio (PSNR) and mean squared error represented performance measures that the results of this study depend on it. The results of various bases of wavelet such as: Daubechies (db), biorthogonal (bior.) and symlet (sym.), show that denoising process that uses in this method produces extra prominent images and improved values of PSNR than other methods.

## 1. INTRODUCTION

Efficient underwater image denoising is a critical aspect for many applications [1]. Underwater images present two main problems: light scattering that alters light path direction and color change. The basic processes in underwater light propagation are scattering and absorption. Underwater noise generally originates from man-made (e.g. shipping and machinery sounds) and natural (e.g. wind, seismic and rain) sources. Underwater noise reduces image quality [1, 2], and denoising has to be applied to improve it [3]. Underwater sound attenuation is dependent on frequency. Consequently, power spectral density (PSD) for ambient noise is defined as colored [4].

Many image denoising techniques are described in [5-9]. A method based on adaptive wavelet with adaptive threshold selection was suggested in [5] to overcome the underwater image denoising problem. Assume that an underwater image has a small signal-to-noise ratio (SNR) and image quality is poor. The simulation results show that the proposed method successfully eliminates noise, improves the peak SNR (PSNR) output of the image and produces a high-quality image. Light is repeatedly deflected and reflected by existing particles in the water due to the light scattering phenomenon, which degrades the visibility and contrast of underwater images. Therefore, underwater images exhibit poor quality. To process images further, wavelet transform and Weber's law were proposed in [8]. Firstly, several pre-processing methodologies were conducted prior to wavelet denoising thresholding. Then, Weber's law was used for image enhancement along with wavelet transform. Consequently, the recovered images were enhanced and the noise level was reduced. In the current study, a novel image denoising method is proposed in the presence of underwater noise using a pre-whitening filter and discrete wavelet transform (DWT) with single-level estimation.

## 2. Ambient Noise Characteristics

The characteristics of underwater noise in seas have been discussed extensively [10]. Such noise has four components: turbulence, shipping, wind and thermal noises. Each component occupies a certain frequency band of spectrum. The PSD of each component is expressed as [11-13].

$$N_t(f) = 17 - 30 \log f . \tag{1}$$

$$N_s(f) = 40 + 20(s - 5) + 26 \log f - 60 \log(f + 0.03) . \tag{2}$$

$$N_w(f) = 50 + 7.5w^{1/2} + 20 \log f - 40 \log(f + 0.4). \tag{3}$$

$$N_{th}(f) = -15 + 20 \log f . \tag{4}$$

where f represents the frequency in KHz. Therefore, the total PSD of underwater noise for a given frequency f (kHz) is

$$Sxx(f) = Nt(f) + Ns(f) + Nw(f) + Nth(f) . \tag{5}$$

Each noise source is dominant in certain frequency bands, as indicated in Table 1.

**Table 1** Underwater noise band

| Band | Type |
|------|------|
| 0.1Hz - 10Hz | Turbulence noise |
| 10Hz - 200Hz | Shipping noise |
| 0.2 kHz - 100 kHz | Wind noise |
| above 100 kHz | Thermal noise |

## 3. Image Model in Presence of Colored Noise

Noise interference is a common problem in digital communication and image processing. An underwater noise model for image denoising in an additive coloured noise channel is presented in this section.

Numerous applications assume that a received image can be expressed as follows:

$$x[n] = s[n] + v[n] \tag{6}$$

where $s(n)$ is the original image and $v(n)$ denotes underwater noise. Hence, denoising aims to eliminate the corruption degree of $s(n)$ caused by $v(n)$. The power spectrum and autocorrelation of additive white Gaussian noise (AWGN) are expressed as [14]:

$$R_{vv}[m] = E\{v[m]v[m + k]\} = \sigma_v^2 \delta[m] \tag{7}$$

$$S_{vv}[e^{j2\pi f}] = F.T._{m \to f}\{R_{vv}[k]\} = \sigma_v^2 \qquad \frac{-f_s}{2} \leq f \leq \frac{f_s}{2} \tag{8}$$

The PSD of AWGN remains constant across the entire frequency range, in which all ranges of frequencies have a magnitude of $\sigma_v^2$. The probability distribution function $\rho_v(v)$ for AWGN is specified by [15]

$$\rho_v(\nu) = \frac{1}{\sigma_v\sqrt{2\pi}} e^{-\frac{\nu^2}{2\sigma_v^2}} \qquad\qquad\qquad (9)$$

where $\sigma_v$ represents the standard deviation. With regard to autocorrelation functions, the delta function indicates that adjacent samples are independent. Therefore, observed samples are considered independent and identically distributed. Underwater noise is dependent on frequency [16, 17]; and it is suitably modelled as colored noise [1, 2, 18]. The PSD of coloured noise is defined as [19, 20]

$$S_{VV}(e^{j2\pi f}) = \frac{1}{f^\beta} \qquad \beta > 0, \qquad \frac{-f_s}{2} \le f \le \frac{f_s}{2} \qquad (10)$$

However, the $R_{vv}[m]$ of coloured noise is not like a delta function, but, it is takes the formula of a $sinc(\ )$ function [14, 19]. In contrast to AWGN, noise samples are correlated [20].

## 4. Image Denoising

Wavelets are used in image processing for sample edge detection, watermarking, compression, denoising and coding of interesting features for subsequent classification [21, 22]. The following subsections discuss image denoising by thresholding the DWT coefficients.

### 4.1 DWT of an image data

An image is presented as a 2D array of coefficients. Each coefficient represents the brightness degree atthat point. Most herbal photographs exhibit smooth colouration variation swith excellent details represented as sharp edges among easy versions. Clean variations in colouration can be strictly labelled as low-frequency versions, whereas pointy variations can be labelled as excessive-frequency versions. The low-frequency components (i.e. smooth versions) establish the base of a photograph, whereas the excessive-frequency components (i.e. the edges that provide the details) are uploaded upon the low-frequency components to refine the image, thereby producing an in-depth image. Therefore, the easy versions are more important than the details. Numerous methods can be used to distinguish between easy variations and photograph information. One example of these methods is picture decomposition via DWT remodelling. The different decomposition levels of DWT are shown in Figure 1.

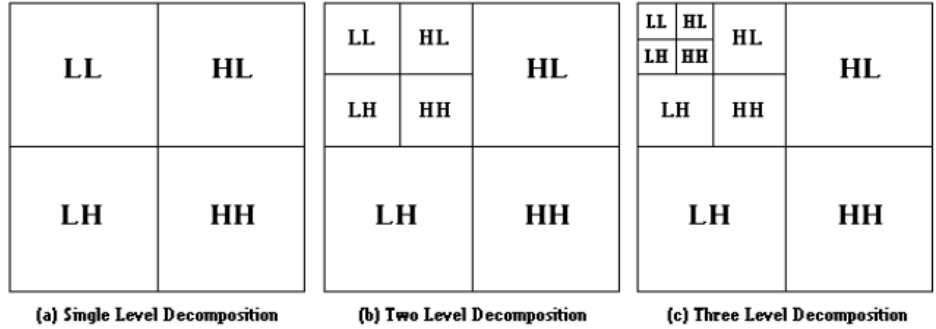

Figure.1 DWT Decomposition levels

## 4.2 The Inverse DWT of an image

Different classes of data are collected into a reconstructed image by using reverse wavelet transform. A pair of high- and low-pass filters is also used during the reconstruction process. This pair of filters is referred to as the synthesis filter pair. The filtering procedure is simply the opposite of transformation; that is, the procedure starts from the highest level. The filters are firstly applied column-wise and then row-wise level by level until the lowest level is reached.

## 5. Proposed Method

In this paper, the DWT is used for the transformation of image in the process of denoising. The advantages of used multi level threshold estimation in denoising process to reduce the required of the use of the prewhitening stage in case of using single level threshold estimation [23]. The following steps describe the image denoising procedure.

1. The DWT of a noisy image is computed. The WT is (time-frequency distribution) that used to decompose signal into family of functions localize in frequency and time. The CWT can be described as:

$$X(t,a) = \frac{1}{\sqrt{a}} \int_{-\infty}^{\infty} x(\tau) h\left(\frac{\tau - t}{a}\right) d\tau \qquad (11)$$

where $t$ is shifting in time and a is scale factor or dilation factor and h(t) is represent basis function. Debauchies, Coiflet, Symlet, and Biorthogonal represent examples of functions used in CWT as shown in Figure 2.

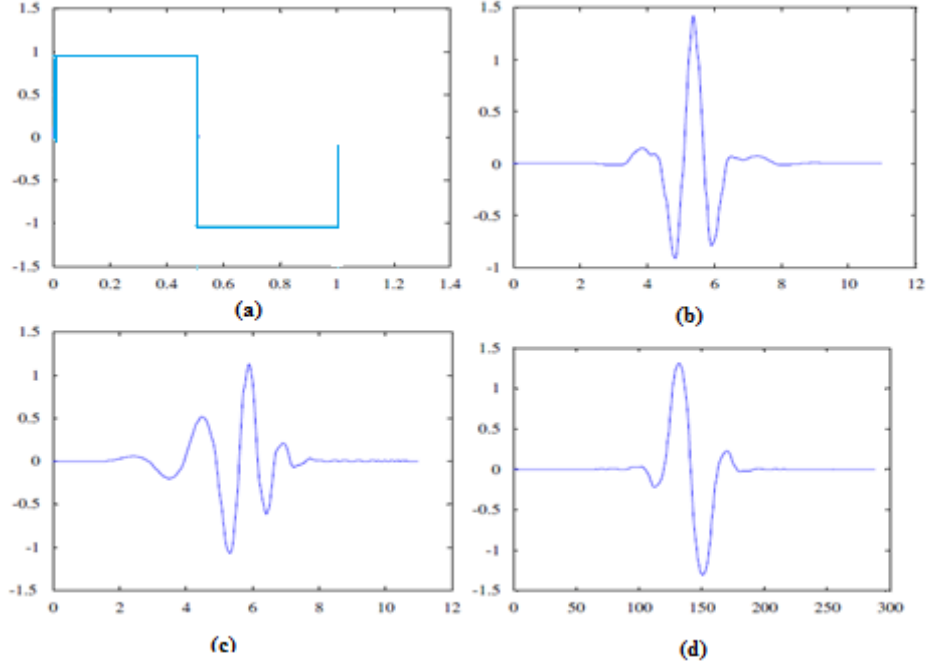

Figure.2 Some basis functions used in WT are: (a) Haar (b) Symlet 6 (c) Debauchies 6 (d) Biorthogonal 1.5 [24]

2. After the DWT representation done, de-noising is done using soft - thresholding by modified universal threshold estimation (MUTE). Providing ambient noise is a colored, a threshold dependent on level applied to each level of  frequency  was proposed in [25]. The value  of threshold applied to the coefficients of   estimated time–frequency using MUTE [25] is expressed as

$$\lambda_k = c . \sigma_{v,k} \sqrt{2\log{(N)}} \qquad (12)$$

where N is length  of signal, $\sigma_{v,k}$ is noise estimated standard deviation for level k, and c is the (modified universal threshold factor) $0 < c < 1$ . The standard deviation for noise at each level is:

$$\sigma_{v,k} = \frac{median(|X(n,k)|)}{0.6745} \qquad (13)$$

where $X(n,k)$ represents all the coefficients for frequency level k [26].

The value of threshold $\lambda_k$ is used to removing the noise and also for efficient recover original signal. The threshold factor c is used to improve further performance of denoising [27]. The value of c is calcualed gradually by increment it of 0.1 for each level to obtain the best results at highest PSNR.

3. after values of threshold $\lambda_k$   is determined for all components, the components representations of time–frequency after hard - thresholding are

$$X_\lambda(n,k) = \begin{cases} X(n,k) & if\, |X(n,k)| > \lambda_k \\ 0 & if\, |X(n,k)| \le \lambda_k \end{cases}$$

(14)

and the components after soft - thresholding are

$$X_\lambda(n,k) = \begin{cases} sgn(X(n,k))\big(|X(n,k)| - \gamma_{k,re}\big) & if\, |X(n,k)| > \lambda_k \\ 0 & if\, |X(n,k)| \le \lambda_k \end{cases}$$

((15)

where $\gamma_k$ denotes the threshold value in level k.

4. The image is reconstructed by applying inverse DWT to obtain the denoised image. The IWT is expressed as:

$$x(\tau) = \int_0^\infty \int_{-\infty}^\infty X(t,a)\frac{1}{\sqrt{a}}h\left(\frac{\tau - t}{a}\right)d\tau\frac{da}{a^2}$$

(16)

Figure 3 shows the data flow diagram of the image denoising process.

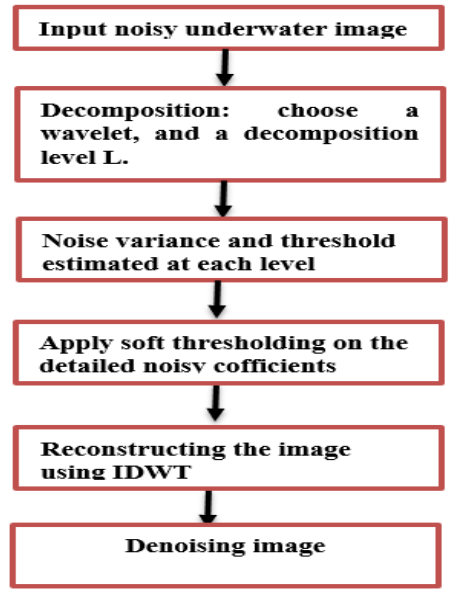

Figure.3 Data flow diagram of image denoising using Level-Dependent Estimation Discrete Wavelet Transform.

## 6. Performance Measures

Common measurement parameters for image reliability include mean absolute error, normalized MSE (NMSE), PSNR and MSE [28]. An SNR over 40 dB provides excellent image quality that is close to that of the original image; an SNR of 30–40 dB typically produces good image quality with acceptable distortion; an SNR of 20–30 dB presents poor image quality; an

SNR below 20 dB generates an unacceptable image [29]. The calculation methods of PSNR and NMSE [30] are presented as follows:

$$PSNR = 10 \log_{10} \frac{255^2}{MSE} \tag{20}$$

where MSE is the MSE between the original image ($x$) and the denoised image ($\hat{x}$) with size M × N:

$$MSE = \frac{1}{M * N} \sum_{i=1}^{M} \sum_{j=1}^{N} [x(i,j) - \hat{x}(i,j)]^2 \tag{21}$$

## 7. Results and Discussion

MATLAB is used as the experimental tool for simulation, and simulation experiments are performed on a diver image to confirm the validity of the algorithm. The simulations are achieved at PSNR ranging from 30 dB to 60 dB by changing noise power from 0 dB to 15 dB. The applied order of the whitening filter is 10. Different denoising wavelet biases (i.e. Debauchies, biorthogonal 1.5 and symlet) are tested on an image with underwater noise via numerical simulation. As shown in Figure 4, soft thresholding and four decomposition levels are used.

| Biaes type | Noisy image | De-noise image | PSNR (dB) |
|---|---|---|---|
| Sym4 | 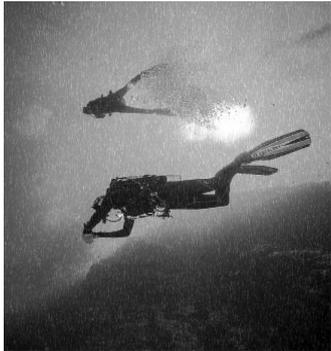 | 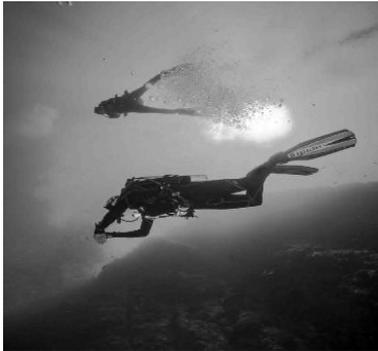 | 45.7 dB |
| Sym4 | 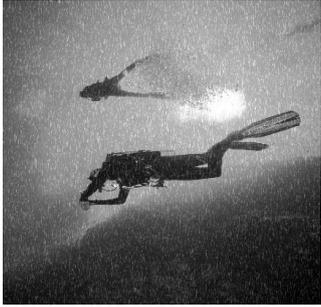 | 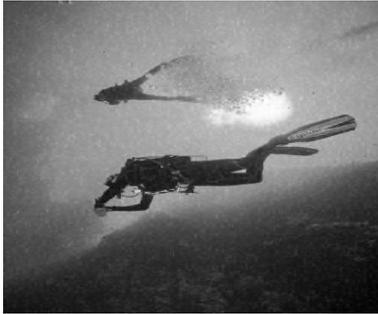 | 29.01dB |
| **db**5 | 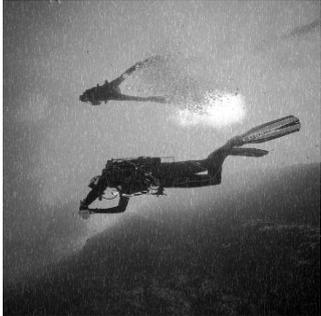 | 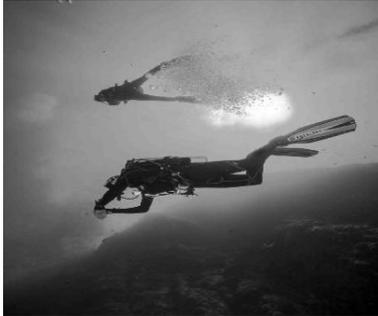 | 45.77 dB |
| **db**5 | 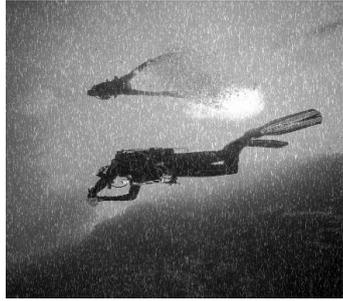 | 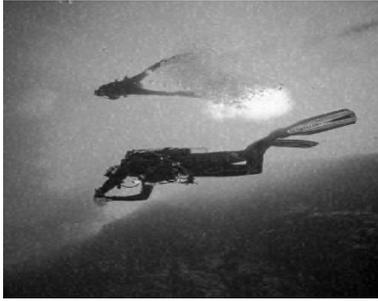 | 28.96 dB |
| Biort hogonal 1.3 | 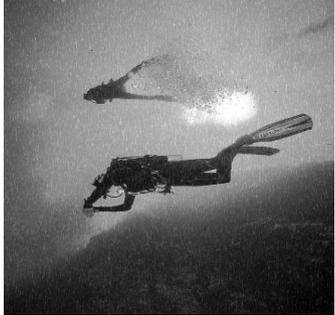 | 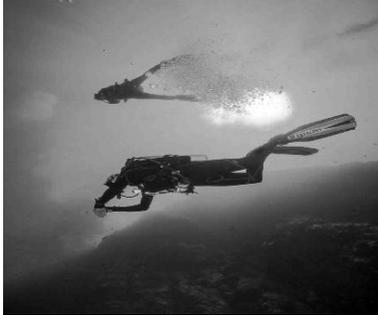 | 45.27 dB |

| Biort hogonal 1.3 | 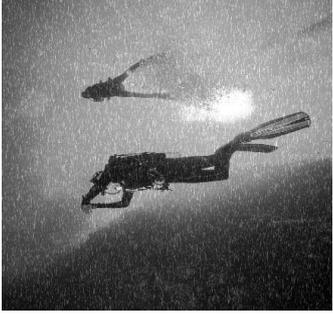 | 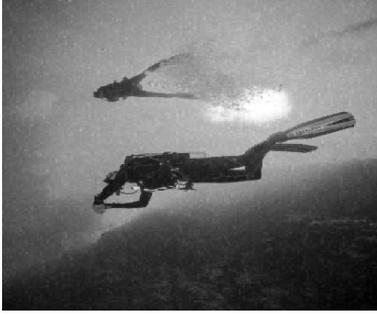 | 28.97dB |
| --- | --- | --- | --- |

Figure.4 Simulation results on diver image using different wavelet biases.

Tables 2, 3 and 4 show the performance of the proposed method on various noise power based on the Debauchies, symlet and biorthogonal wavelet biases, respectively. The PSNR and MSE values are calculated based on each noise power value.

Table 2 Performance results of PSNR and MSE on diver image based on Debauchies wavelet bias

| Noise power (db) | PSNR | MSE |
| --- | --- | --- |
| 0 | 59.35 | 0.0756 |
| 3 | 55.57 | 0.2161 |
| 5 | 52.3 | 0.5226 |
| 10 | 42.9653 | 0.5062 |
| 15 | 33.2530 | 0.8279 |

Table 3 Performance results of PSNR and MSE on diver image based on symlet wavelet bias

| Noise power (db) | PSNR | MSE |
| --- | --- | --- |
| 0 | 61.066 | 0.0759 |
| 3 | 55.48 | 0.2948 |
| 5 | 52.032 | 0.5509 |
| 10 | 43.33 | 0.4420 |
| 15 | 35.0591 | 0.392 |

Table 4 Performance results of PSNR and MSE on diver image based on biorthogonal wavelet bias

| Noise power (db) | PSNR | MSE |
| --- | --- | --- |
| 0 | 59.7560 | 0.0773 |
| 3 | 55.76 | 0.321 |
| 5 | 52.867 | 0.5240 |
| 10 | 43.7970 | 0.767 |
| 15 | 35.125 | 0.8604 |

## 8. Conclusion

Underwater noise is mainly characterised as non-white and non-Gaussian noise. Therefore, traditional methods used for image denoising using wavelet transform underwater are inefficient because these methods use multi - level estimation discrete wavelet transform for noise variance. However, noise variance at each level should be independently estimated in coloured noise. The wavelet denoising method can be efficiently used with PSNR and MSE compared to other method used a pre-whitening filter that converts underwater noise to white noise, as demonstrated by the results.

**REFRENCES**: